\documentclass[twoside,twocolumn,9pt]{article}
\usepackage{extsizes}
\usepackage[super,sort&compress,comma]{natbib} 
\usepackage[version=3]{mhchem}
\usepackage[left=1.5cm, right=1.5cm, top=1.785cm, bottom=2.0cm]{geometry}
\usepackage{balance}
\usepackage{times,mathptmx}
\usepackage{amsmath,amssymb,amsfonts}
\usepackage{dsfont}
\usepackage{sectsty}
\usepackage{graphicx} 
\usepackage{lastpage}
\usepackage[format=plain,justification=justified,singlelinecheck=false,font={stretch=1.125,small,sf},labelfont=bf,labelsep=space]{caption}
\usepackage{float}
\usepackage{fancyhdr}
\usepackage{fnpos}
\usepackage[english]{babel}
\usepackage{array}
\usepackage{droidsans}
\usepackage{charter}
\usepackage[T1]{fontenc}
\usepackage[usenames,dvipsnames]{xcolor}
\usepackage{setspace}
\usepackage[compact]{titlesec}
\usepackage{placeins}
\usepackage{braket}
\usepackage{hyperref}

\newcommand*{\Laplace}{\boldsymbol{\triangle}}%
\newcommand*{\dif}{\mathrm{d}}%
\newcommand*{\norm}[1]{\lVert#1\rVert}%
\newcommand*{\Eins}{\mathds{1}}%
\newcommand*{\Nabla}{\vec{\nabla}}%
\newcommand*{\onlinecite}[1]{[\hspace{-1ex} \nocite{#1}\citenum{#1}]}%

\allowdisplaybreaks
\definecolor{cream}{RGB}{222,217,201}

\begin{document}

\pagestyle{fancy}
\thispagestyle{plain}
\fancypagestyle{plain}{%
\renewcommand{\headrulewidth}{0pt}}

\makeFNbottom
\makeatletter
\renewcommand\LARGE{\@setfontsize\LARGE{15pt}{17}}
\renewcommand\Large{\@setfontsize\Large{12pt}{14}}
\renewcommand\large{\@setfontsize\large{10pt}{12}}
\renewcommand\footnotesize{\@setfontsize\footnotesize{7pt}{10}}
\makeatother

\renewcommand{\thefootnote}{\fnsymbol{footnote}}
\renewcommand\footnoterule{\vspace*{1pt}%
\color{cream}\hrule width 3.5in height 0.4pt \color{black}\vspace*{5pt}} 
\setcounter{secnumdepth}{5}

\makeatletter 
\renewcommand\@biblabel[1]{#1}            
\renewcommand\@makefntext[1]%
{\noindent\makebox[0pt][r]{\@thefnmark\,}#1}
\makeatother 
\renewcommand{\figurename}{\small{Fig.}~}
\sectionfont{\sffamily\Large}
\subsectionfont{\normalsize}
\subsubsectionfont{\bf}
\setstretch{1.125} 
\setlength{\skip\footins}{0.8cm}
\setlength{\footnotesep}{0.25cm}
\setlength{\jot}{10pt}
\titlespacing*{\section}{0pt}{4pt}{4pt}
\titlespacing*{\subsection}{0pt}{15pt}{1pt}

\fancyfoot{}
\fancyfoot[RO]{\footnotesize{\sffamily{\thepage}}}
\fancyfoot[LE]{\footnotesize{\sffamily{\thepage}}}
\fancyhead{}
\renewcommand{\headrulewidth}{0pt} 
\renewcommand{\footrulewidth}{0pt}
\setlength{\arrayrulewidth}{1pt}
\setlength{\columnsep}{6.5mm}
\setlength\bibsep{1pt}

\makeatletter 
\newlength{\figrulesep} 
\setlength{\figrulesep}{0.5\textfloatsep} 
\newcommand{\topfigrule}{\vspace*{-1pt}\noindent{\color{cream}\rule[-\figrulesep]{\columnwidth}{1.5pt}} }
\newcommand{\botfigrule}{\vspace*{-2pt}\noindent{\color{cream}\rule[\figrulesep]{\columnwidth}{1.5pt}} }
\newcommand{\dblfigrule}{\vspace*{-1pt}\noindent{\color{cream}\rule[-\figrulesep]{\textwidth}{1.5pt}} }
\makeatother

\twocolumn[
\begin{@twocolumnfalse}
\vspace{3cm}
\sffamily
\begin{tabular}{m{4.5cm} p{13.5cm} }
&\noindent\LARGE{\textbf{Analytical approach to chiral active systems: suppressed phase separation of interacting Brownian circle swimmers}} \\
\vspace{0.3cm} & \vspace{0.3cm} \\
&\noindent\large{Jens Bickmann \textit{$^{a}$}, 
Stephan Br\"oker \textit{$^{a}$},
Julian Jeggle \textit{$^{a}$},
and Raphael Wittkowski \textit{$^{a}$}}$^{\ast}$ \\
&\noindent\normalsize{We consider chirality in active systems by exemplarily studying the phase behavior of planar systems of interacting Brownian circle swimmers with a spherical shape. Continuing previous work presented in [G.-J.\ Liao, S.\ H.\ L.\ Klapp, \textit{Soft Matter}, 2018, \textbf{14}, 7873-7882], we derive a predictive field theory that is able to describe the collective dynamics of circle swimmers. The theory yields a mapping between circle swimmers and noncircling active Brownian particles and predicts that the angular propulsion of the particles leads to a suppression of their motility-induced phase separation, being in line with previous simulation results. In addition, the theory provides analytical expressions for the spinodal corresponding to the onset of motility-induced phase separation and the associated critical point as well as for their dependence on the angular propulsion of the circle swimmers. We confirm our findings by Brownian dynamics simulations and an analysis of the collective dynamics using a weighted graph-theoretical network. The agreement between results from theory and simulation is found to be good. 
}\\
\end{tabular}
\end{@twocolumnfalse}\vspace{0.6cm}]

\renewcommand*\rmdefault{bch}\normalfont\upshape
\rmfamily
\section*{}
\vspace{-1cm}

\footnotetext{\textit{$^{a}$~Institut f\"ur Theoretische Physik, Center for Soft Nanoscience, Westf\"alische Wilhelms-Universit\"at M\"unster, D-48149 M\"unster, Germany}}
\footnotetext{\textit{$^{\ast}$~Corresponding author: raphael.wittkowski@uni-muenster.de}}

\section{Introduction}
Active systems can show a variety of collective behaviors, including phenomena like shoaling and schooling of fish \cite{Ballerini_2008,Becker_2015}, flocking and swarming of birds \cite{Bialek_2012,Qiu_2015,Liebchen2017}, cell migration \cite{Liu2011, Liu2017}, swirling \cite{Dombrowski_2004,Riedel_2005,Kudrolli_2008}, laning \cite{Vissers_2011,Kogler_2015,W_chtler_2016}, low-Reynolds-number turbulence \cite{Heidenreich_2016,Thampi_2016t,Thampi_2016t2,Doostmohammadi_2017,Kokot_2017,Reinken_2018, Reichhardt_2018}, clustering \cite{Buttinoni_2013,Bialk__2015,Lavrentovich_2016,Navarro_2015,Ishimoto_2018,Reichhardt_2018,Speck_2016}, crystallization \cite{Kumar_2018,Praetorius_2018}, superfluidity \cite{Marchetti_2015}, and even the emergence of negative viscosities \cite{Saintillan_2018}. The active agents can be macroscopic like pedestrians \cite{Helbing_1995,AppertRolland18} or animals \cite{Ballerini_2008,Bialek_2012,Becker_2015,Qiu_2015} but also microscopic like bacteria or colloidal Janus particles \cite{waltherM2008janus,waltherM2013janus,ebbensG2018catalytic,campbellEIG2019experimental}.

In the literature, usually particles with symmetric active propulsion and thus no chirality or helicity are considered. However, most systems are imperfect leading to, e.g., chirality in a two-dimensional geometry \cite{bechinger2016active,Lwen2016}. Since this generates a circular motion, these particles are referred to as \textit{circle swimmers} \cite{Teeffelen08}. They are known already since 1901 from studies regarding microorganisms by Jennings \cite{Jennings1901}. Chirality and chiral motion can originate from different types of asymmetry. A common example is a shape asymmetry leading to an effective propulsion torque, as it is present, e.g., for an `L'-shaped particle \cite{K_mmel_2013,kummeltHWTBEVLB2014reply}. Further examples are Janus particles with an asymmetric coating \cite{Mano17}, bacteria with unequally strong flagella \cite{Brokaw1982,Kamiya1984}, and structure formation of achiral particles \cite{Zhang16}. Even external fields (such as, e.g., gravitational \cite{Campbell17}, magnetic \cite{Schirmacher15}, or temperature \cite{Ai19} fields) and nearby surfaces \cite{Daddi18} can bend trajectories. Especially the chiral behavior at walls plays an important role in nature and was studied for \textit{Escherichia coli} bacteria \cite{DiLuzio_2005} as well as spermatozoa \cite{Nosrati_2017}. The latter's particle interaction at walls is known to play a crucial role in the fertilization process \cite{Nosrati_2015}.

Individual circle swimmers were studied theoretically \cite{Teeffelen08,ten2009non,vanTeeffelen2009,Mijalkov2013,Volpe2014,Schirmacher15,Campbell17,Daddi18,Markovich2019} as well as experimentally \cite{Brokaw1982,Kamiya1984,DiLuzio_2005,K_mmel_2013,kummeltHWTBEVLB2014reply,Nosrati_2015,Campbell17,Mano17,ebbensG2018catalytic}. There exist some works on alignment of circle swimmers via anisotropic particle shapes \cite{Denk16,Baer19} and an imposed alignment mechanism as in the Vicsek model \cite{Vicsek1995, Liebchen2017, Levis19,LevisPL2019activity,Barre2015} as well as a description that models particle interaction via coupling to a field describing signalling molecules \cite{Liebchen16}. The collective behavior of circle swimmers, however, is not yet fully understood. A prime example for the collective behavior of active colloidal particles is motility-induced phase separation (MIPS) \cite{Cates15mips,Speck_2016}. MIPS is the effect of forming a dense liquid-like phase of particles and a dilute gas-like phase of particles out of a homogeneous distribution of repulsively interacting active particles. However, studies of MIPS in systems of circle swimmers are rare. There exists a work of Liao and Klapp \cite{Liao2018} investigating MIPS in systems of circle swimmers by computer simulations and an investigation of MIPS in circle-swimmer systems at vanishing temperature on the basis of computer simulations and dynamic mean-field theory by Lei, Ciamarra, and Ni \cite{Lei19}. Another simulation study on the collective dynamics of circle swimmers showed that chiral active particles can be rectified by transversal temperature differences \cite{Ai19}. In addition to these mostly computer-simulation-based studies, a theory for rod-like active Brownian particles (ABPs) was recently published \cite{vanDamme19}. It suggests a suppression of MIPS due to chirality. The authors also discussed the idea of an effective rotational diffusion. Their work, however, describes the dynamics by effective quantities and provides no analytical prediction or result for the effective rotational diffusion or for the spinodal corresponding to the onset of MIPS.
There exists also a dynamical density functional theory (DDFT) \cite{teVrugtLW2020DDFTreview} for circle swimmers \cite{Hoell_2017}, but it cannot predict the spinodal for MIPS, since the DDFT approach is limited to low particle densities and weak propulsion, where MIPS does not occur \cite{teVrugtLW2020DDFTreview}. However, there is recent work on a predictive hydrodynamic theory for mixtures of circle swimmers \cite{LevisPL2019activity} that is based on the Vicsek model. This hydrodynamic theory can describe effects like flocking \cite{Liebchen2017}, but it is unable to describe clustering and MIPS, since no spatial interactions are taken into account. This problem will be addressed in the present work.

We study ABPs as an important class of active colloidal particles, where hydrodynamic interactions are not present but the hydrodynamic resistance exerted on the particles is taken into account, and consider a translational and rotational propulsion of these particles leading to circle swimming.
Extending previous predictive field-theoretical models for ABPs \cite{BickmannW2020, BickmannW2020b} towards circle swimmers, we obtain a $2$nd-order-derivatives model that describes the collective dynamics of circle swimmers. Interestingly, we find that the diffusive dynamics of circle swimmers can be mapped onto that of ABPs via an effective rotational diffusion coefficient. Furthermore, this model predicts a spinodal that we compare for different values of the particles' angular propulsion velocity to data we obtained from Brownian dynamics simulations. The critical point associated with the spinodal is found to have a remarkable dependence on the angular propulsion velocity. While the critical density is independent of the angular propulsion velocity, the critical P\'eclet number describing the activity of the particles at the critical point increases with the angular propulsion velocity.

This article is structured as follows. In section \ref{sec:Methods}, we derive the analytical model and present the details of our computer simulations. We discuss our results in section \ref{sec:Phasebehavior}. Finally, we conclude in section \ref{sec:Conclusions}.

\section{\label{sec:Methods}Methods}
\subsection{\label{ssec:AnalyticalModel}Analytical model}
We consider a suspension of $N$ spherical Brownian circle swimmers in two spatial dimensions. Each particle has a constant translational propulsion parallel to its instantaneous orientation and an additional constant angular propulsion. We denote the translational speed of a free particle by $v_0$ and the angular velocity by $\omega$. The center of mass of the $i$-th particle is denoted by the vector $\vec{r}_i = (x_1, x_2)^\mathrm{T}$ and its orientation by the unit vector $\hat{u}(\phi_i) = (\cos(\phi_i), \sin(\phi_i))^\mathrm{T}$ that is parametrized by the polar angle $\phi_i$. $\vec{r}_i(t)$ and $\phi_i(t)$ depend on time $t$. 
The dynamics of the particles can be described via the overdamped Langevin equations for circle swimmers \cite{vanTeeffelen2009,Mijalkov2013,Volpe2014,bechinger2016active,Liao2018,Ai19}
\begin{align}
\dot{\vec{r}}_i &= v_0\hat{u}(\phi_i) + \beta D_\mathrm{T} \vec{F}_{\mathrm{int}, i}(\{\vec{r}_i\})
+ \vec{\xi}_{\mathrm{T},i}, \label{eqn:LangevinR}\\
\dot\phi_i &= \omega + \xi_{\mathrm{R},i}, \label{eqn:LangevinPHI}%
\end{align}
where a dot over a variable denotes a derivative with respect to time, 
$\beta =1/(k_{\mathrm{B}}T)$ is the thermodynamic beta with the Boltzmann constant $k_{\mathrm{B}}$ and absolute temperature $T$,
$D_\mathrm{T}$ is the translational diffusion coefficient of the particles, 
$\vec{F}_{\mathrm{int}, i}(\{\vec{r}_i\})$ describes the interaction force acting on the $i$-th particle, and $\vec{\xi}_{\mathrm{T}, i}(t)$ and $\xi_{\mathrm{R}, i}(t)$ are statistically independent Gaussian white noises with zero mean for the translational and rotational degrees of freedom of the $i$-th particle, respectively. The correlations of the noise terms are given by $\braket{\vec{\xi}_{\mathrm{T}, i}(t_1)\otimes\vec{\xi}_{\mathrm{T}, j}(t_2)} = 2D_\mathrm{T}\Eins_2 \delta_{ij}\delta(t_1-t_2)$ and $\braket{\xi_{\mathrm{R}, i}(t_1)\xi_{\mathrm{R}, j}(t_2)} = 2D_\mathrm{R}\delta_{ij}\delta(t_1-t_2)$ with the ensemble average $\langle\cdot\rangle$, dyadic product $\otimes$, $2\times 2$-dimensional identity matrix $\Eins_2$, Kronecker delta $\delta_{ij}$, Dirac delta function $\delta(t)$, and rotational diffusion constant $D_{\mathrm{R}}$. The angular velocity $\omega$ constitutes the difference between ABPs in two spatial dimensions \cite{Stenhammar2014,Speck2016,RW,Digregorio2018,Duzgun2018,Siebert2018,Tjhung2018, BickmannW2020, Jeggle2019} and the circle swimmers considered here and in Ref.\ \onlinecite{Liao2018}\footnote{The angular velocity $\omega$ introduced here is equivalent to $\omega_0$ in Ref.\ \onlinecite{Liao2018}.}. As usual, we consider an additive pairwise interaction so that the interaction force can be written as $\vec{F}_{\mathrm{int},i} = -\sum_{j=1, j\neq i}^{N}\Nabla_{\vec{r}_i} U_2(\norm{\vec{r}_i-\vec{r}_j})$, where $\Nabla_{\vec{r}_i} = (\partial_{x_{1, i}}, \partial_{x_{2, i}})^\mathrm{T}$ is the del operator in two spatial dimensions with respect to $\vec{r}_i$ and $U_2(\norm{\vec{r}_i-\vec{r}_j})$ is the pair-interaction potential depending on the distance of the $i$-th and $j$-th particles. In the following, we derive a predictive local field-theoretical model for the collective dynamics of the circle swimmers based on the Langevin equations \eqref{eqn:LangevinR} and \eqref{eqn:LangevinPHI}.

The statistically equivalent Smoluchowski equation corresponding to the Langevin equations \eqref{eqn:LangevinR}and \eqref{eqn:LangevinPHI} is given by
\begin{equation} 
\begin{split}
\dot{P} & = \sum_{i=1}^N (D_\mathrm{T}\Laplace_{\vec{r}_i} + D_\mathrm{R}\partial_{\phi_i}^2 -\omega\partial_{\phi_i})P \\
&\qquad\,\,\, - \Nabla_{\vec{r}_i}\cdot \big( ( \beta D_\mathrm{T}\vec{F}_{\mathrm{int}, i}(\{\vec{r}_i\}) + v_0\hat{u}(\phi_i) ) P\big).\label{eqn:SMOLUCHOWKSI}
\end{split}
\end{equation}
Here, $P$ denotes the many-particle probability density $P(\lbrace  \vec{r}_i\rbrace, \lbrace \phi_i \rbrace, t)$ and $\Laplace_{\vec{r}_i}$ is the Laplacian with respect to $\vec{r}_i$. 
At this point in the derivation, the angular velocity gives rise to the term $-\omega\partial_{\phi_i}P$, which is of first order in the angular derivative. Remarkably, there is no such term in corresponding models for ordinary ABPs in two \cite{BickmannW2020} and three \cite{BickmannW2020b} spatial dimensions. As we will show further below, this term has important consequences.

\begin{figure}[ht]
\centering
\includegraphics[width=8.64cm]{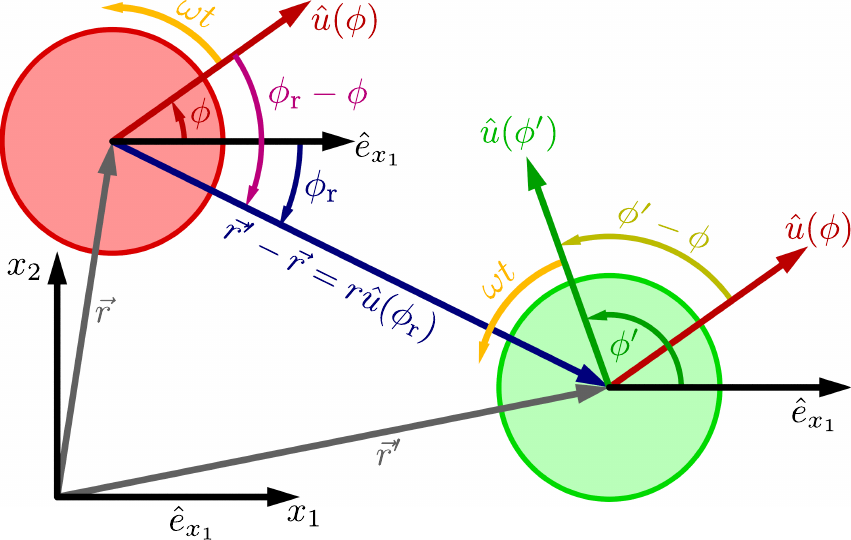}
\caption{\label{fig:geometry}Absolute and relative positions and orientations of two circle swimmers with angular velocity $\omega$, where $\hat{e}_{x_1}$ denotes the unit vector in $x_1$ direction.}
\end{figure}

We derive our model for the collective dynamics of circle swimmers via the \textit{interaction-expansion method} \cite{RW,BickmannW2020, BickmannW2020b}. As a first step, we calculate the one-particle density $\psi$ by integrating both sides of the Smoluchowski equation over the degrees of freedom of all but one particles and renaming its position vector and orientation angle as $\vec{r}$ and $\phi$, respectively:
\begin{equation}
\psi(\vec{r}, \phi, t) = N \bigg(\prod_{\begin{subarray}{c}j = 1\\j\neq i\end{subarray}}^{N} \int_{\mathbb{R}^2}\!\!\!\!\dif^{2}r_j\int_{0}^{2\pi}\!\!\!\!\!\!\!\dif\phi_j \bigg)\,  P\bigg\rvert_{\begin{subarray}{l}\vec{r}_i=\vec{r},\\\phi_i=\phi\end{subarray}}.
\label{eqn:OPD}%
\end{equation}
Following the procedure described in Refs.\ \onlinecite{BickmannW2020,BickmannW2020b}, we perform a Fourier expansion, a gradient expansion \cite{Yang1976,Evans1979,EmmerichLWGTTG12}, a Cartesian orientational expansion \cite{TeVrugt19}, and a quasi-stationary approximation \cite{RW,BickmannW2020,BickmannW2020b}. Moreover, the pair-distribution function $g(\vec{r}, \vec{r}', \phi, \phi', t)$ that relates the two-particle density $\psi^{(2)}(\vec{r}, \vec{r}', \phi, \phi', t)$ to the one-particle density $\psi(\vec{r}, \phi, t)$ is introduced:
\begin{equation}
\psi^{(2)}(\vec{r}, \vec{r}', \phi, \phi', t) = g(\vec{r}, \vec{r}', \phi, \phi', t) \psi(\vec{r}, \phi, t)\psi(\vec{r}', \phi', t).
\end{equation}
By assuming translational and rotational invariance as well as time invariance of the system, which correspond to a homogeneous stationary state, the dependence of the pair-distribution function reduces to $g(r, \phi_\mathrm{r}-\phi, \phi'-\phi)$ with the parametrization $\vec{r}'-\vec{r} = r\hat{u}(\phi_\mathrm{r})$ of the relative position $\vec{r}'-\vec{r}$. Since $\omega$ leads to a breaking of rotational symmetries, the assumption of rotational invariance is well justified in the limit $\omega = 0$ but becomes increasingly inaccurate for, according to amount, larger angular velocities. A sketch of the setup illustrating the used absolute and relative coordinates of two particles is shown in Fig.\ \ref{fig:geometry}. 

For the dynamics of the local density field $\rho(\vec{r}, t) = \int_0^{2\pi}\dif\phi\, \psi(\vec{r}, \phi, t)$, one then obtains, up to the $2$nd order in derivatives, the model\footnote{From here onwards, summation over indices appearing twice in a term is implied.} 
\begin{equation}
\dot\rho(\vec{r}, t) = \partial_i(D(\rho)\partial_i \rho)
\label{eqn:2ndOrderModel}%
\end{equation}
with the compact notation $\partial_i = (\Nabla)_i$, the density-dependent diffusion coefficient 
\begin{equation}
\begin{split}
D(\rho) &=  D_\mathrm{T} + \frac{A(1, 0,0)}{\pi}\rho \\
&\quad \,\, + \frac{1}{2D_\mathrm{R}(1+\omega^{*2})}\Big(v_0  -\frac{2}{\pi}(A(0, 1,0)+A(0, 1,-\!1))\rho\Big)\\
&\qquad\qquad\qquad\qquad\;\, \Big(v_0 - \frac{4}{\pi} A(0, 1,0)\rho\Big),\label{eqn:diffmatrix}%
\end{split}
\end{equation}
and the dimensionless rescaled angular velocity $\omega^* = \omega/D_\mathrm{R}$\cite{Liao2018}. The coefficients $A(m, k_1, k_2)$ are given by
\begin{equation}
A(m, k_1,k_2) = -\pi^2\beta D_\mathrm{T}\int_{0}^{\infty}\!\!\!\!\!\dif r\, r^{m+1}U_2'(r)g_{k_1, k_2}(r)
\end{equation}
with the shorthand notation $U_2'(r)=\dif U_2(r)/\dif r$ and the expansion coefficients
\begin{equation}
g_{k_1, k_2}(r) = \frac{\int_0^{2\pi}\!\dif t_1\int_0^{2\pi}\!\dif t_2\, g(r, t_1, t_2)
\cos(k_1t_1+k_2t_2)}{\pi^2(1+\delta_{k_1, 0})(1+\delta_{k_2, 0})}.
\end{equation}
Remarkably, the obtained diffusion equation has a scalar diffusion coefficient and, albeit the chirality of the particles, no mixing of spatial derivatives occurs\footnote{Actually, a matrix-like diffusion coefficient and a mixing of spatial derivatives were obtained during the derivation. However, the nondiagonal terms cancel each other out, since the diffusion matrix was built upon two base matrices: the Kronecker delta $\delta_{ij}$ and the later introduced \textit{antisymmetric} two-dimensional Levi-Civita symbol $\epsilon_{ij}$ given by Eq.\ \eqref{epsilonij}. For the special case of a $2$nd-order-derivatives model, no contributions from $\epsilon_{ij}$ can occur due to its properties. For higher-order models, there exist a tensorial diffusion coefficient and mixing of spatial derivatives.}.

The $2$nd-order-derivatives model \eqref{eqn:2ndOrderModel} for circle swimmers can be compared with the model for ABPs in two spatial dimensions presented in Ref.\ \onlinecite{BickmannW2020}. There, the achiral case $\omega = 0$ is considered and a $2$nd-order-derivatives model was derived using analogous approximations. The comparison of that model, given by Eqs.\ (21) and (22) in Ref.\ \onlinecite{BickmannW2020}, with ours yields an interesting and important result: The collective dynamics of circle swimmers are, up to the considered orders, qualitatively identical to the collective dynamics of ordinary ABPs, but the former have a larger rotational diffusion. It is therefore possible to map the dynamics of circle swimmers onto that of ordinary ABPs with an effective rotational diffusion coefficient
\begin{equation}
D_\mathrm{R, eff} = D_{\mathrm{R}}(1+\omega^{*2}).
\label{eqn:DReff}%
\end{equation}
This mapping is similar to the mapping of ABP systems onto detailed-balance systems via an effective temperature $T_{\mathrm{eff}}$ \cite{Howse07,LoiMossa08,TailleurCates08,palacci2010sedimentation,sevilla2015}. 
For two spatial dimensions, the effective temperature is given by \cite{preisler2016configurational}
\begin{equation}
T_\mathrm{eff}=T\bigg(1+\frac{\mathrm{Pe}^2}{6}\bigg)
\label{eqn:Teff}%
\end{equation}
with the P\'eclet number $\mathrm{Pe} = v_0\sigma/D_\mathrm{T}$, where $\sigma$ is the diameter of the particles. 
Indeed, one can define an effective rotational temperature
\begin{equation}
T_\mathrm{R,eff}=T(1+\omega^{*2})
\label{eqn:TReff}%
\end{equation}
that corresponds to the orientational motion of the circle swimmers, where $\omega^{*}$ can be considered as a rotational P\'eclet number. Taking into account that, via the Einstein relation, $D_\mathrm{R}$ is proportional to the temperature $T$ so that $D_\mathrm{R, eff}$ should be proportional to $T_\mathrm{R,eff}$, one can justify the similar scaling of $D_\mathrm{R, eff}$ with $\omega^*$ and of $T_\mathrm{eff}$ with $\mathrm{Pe}$.
We stress that the validity of the mapping we found is not restricted to low-density and near-equilibrium systems, which is the case for the mapping of ABPs onto passive Brownian particles. Reasons for this are the fact that angular interactions are not present in the case of spherical particles as considered here and the fact that our derivation is applicable even for arbitrarily large $\mathrm{Pe}$. However, this mapping does not account for chirality effects and is only moderately applicable for higher-order models. It should therefore be applied predominantly for low values of $\omega^*$. 

If one wants to study the dynamics of the circle swimmers explicitly, the values of the coefficients $A(n, k_1, k_2)$ have to be known. This requires information about the product $U'_2(r)g(r,t_1, t_2)$. For a Weeks-Chandler-Andersen (WCA) interaction potential \cite{Weeks_Chandler_Andersson} 
\begin{equation}U_2(r) = \begin{cases}
4\varepsilon\left(\left(\frac{\sigma}{r}\right)^{12}-\left(\frac{\sigma}{r}\right)^{6}\right)+\varepsilon, &\text{if } r<2^{\frac{1}{6}}\sigma,\\
0, & \text{else},
\end{cases}
\end{equation}
where $r=\norm{\vec{r}}$ is the center-of-mass distance of two particles and $\varepsilon$ is the interaction energy, this product is analytically known in two \cite{Jeggle2019} and three \cite{Broeker2019} spatial dimensions. The WCA interaction potential is purely repulsive and often used for ABP simulations \cite{Stenhammar2014,Liao2018,Siebert2018,Jeggle2019,Broeker2019}, since it suits well the interaction behavior of active particles used in experiments \cite{Buttinoni_2013}. Using the available analytical results for the pair-distribution function, the coefficients occurring in the density-dependent diffusion coefficient \eqref{eqn:diffmatrix} are approximately given by \cite{Jeggle2019,BickmannW2020}
\begin{align}
A(1, 0,0)& = 38.2 + 18.4e^{2.87\Phi},\\
A(0, 1,0)& = 36.9\label{eqn:A010coeff},\\
A(0, 1,-\!1)& = -0.232 - 13.6 \Phi,
\end{align}
where $\Phi = \rho\pi\sigma^2/4$ denotes the overall packing density.

Additionally, for potential further applications, we derive a phase-field model describing the collective dynamics of circle swimmers up to $4$th order in derivatives. Since the model would be rather lengthy and complicated, we make a further approximation by neglecting terms of second or higher order in $\omega$.
The model can be written in conservative form
\begin{equation}
\dot\rho = -\partial_i(J_i^{(\mathrm{chiral})} + J_i^{(\mathrm{achiral})}) \label{eqn:PhaseFieldModel}
\end{equation}
with the model-specific current
\begin{align}
\begin{split}
J_i^{(\mathrm{chiral})} &= \omega\epsilon_{ij}(\mu_{1}\rho+\mu_{2}\rho^2+\mu_{3}\rho^3+\mu_{4}\rho^4)\partial_j\Laplace\rho\\
&\quad\:\! + \omega\epsilon_{ij}(\mu_{5}+\mu_{6}\rho+\mu_{7}\rho^2)(\partial_j\rho)(\partial_k\rho)^2\label{eqn:4thOrderModel}
\end{split}
\end{align}
and the current $J_i^{(\mathrm{achiral})}$ for the achiral case $\omega = 0$, which is given by Eq.\ (25) in Ref.\ \onlinecite{BickmannW2020}. Expressions for the coefficients $\mu_i$ are given in the Appendix.
Considering this order in derivatives, a mixing of spatial derivatives occurs. In two spatial dimensions, this mixing is characterized by the two-dimensional Levi-Civita matrix 
\begin{equation}
\epsilon = \begin{pmatrix}
0 & 1 \\
-1 & 0
\end{pmatrix}\label{epsilonij},
\end{equation}
which obeys the relation $\partial_\phi \hat{u}(\phi) = -\epsilon \hat{u}(\phi)$. In the case of the phase-field model \eqref{eqn:PhaseFieldModel}, terms proportional to $\epsilon_{ij}$ are responsible for the mixing of spatial derivatives.

\subsection{\label{ssec:Simulations}Computer simulations}
We carried out Brownian dynamics simulations by integrating the Langevin equations \eqref{eqn:LangevinR} and \eqref{eqn:LangevinPHI} using a modified version of the molecular dynamics simulation package LAMMPS\cite{Plimpton1995}.
In these simulations, we expressed physical quantities in terms of dimensionless Lennard-Jones units, where $\sigma$, the Lennard-Jones time $\tau_{\mathrm{LJ}} = \sigma^2/(\varepsilon\beta D_\mathrm{T})$, and $\varepsilon$ are used as units for length, time, and energy, respectively. The P\'eclet number was varied only by varying $T$ and we fixed $F_A = 24\varepsilon/\sigma$ as well as $D_\mathrm{T}/(k_B T) = \sigma^2/(\varepsilon\tau_{\mathrm{LJ}})$ analogous to Refs.\ \onlinecite{Stenhammar2014,RW,Jeggle2019,Broeker2019}. Following the Stokes-Einstein-Debye relation for spheres, the rotational diffusion coefficient is given by $D_\mathrm{R} = 3 D_\mathrm{T}/\sigma^2$. As simulation domain, a quadratic area with side length $\ell = 128\sigma$ and periodic boundary conditions was chosen. For the initial condition, we used a random distribution of the particles. Particle trajectories were integrated for a total simulation period of $250 \tau_{\mathrm{LJ}}$ with a time-step size $\Delta t = 5 \cdot 10^{-5} \tau_{\mathrm{LJ}}$. The number of particles we simulated ranged from $4172$ to $18775$, based on the overall packing density $\Phi$ of the system. The latter ranged as $\Phi \in [0.2, 0.9]$. 

To study the collective behavior of circle swimmers including the emergence of MIPS, we calculated state diagrams that show the state of a system as a function of the P\'eclet number and the packing density.
However, this requires a good measure that allows to distinguish MIPS from other states of the system. 
In the following, we propose a method that is based on graph-theoretical networks.
We discarded the initial period of duration $125 \tau_{\mathrm{LJ}}$ from the simulation data to give all considered systems enough time to relax to a stationary state. Afterwards, we extracted the particles' positions every $2.5 \tau_{\mathrm{LJ}}$. 
From these positions, we constructed a graph-theoretical network: The nodes are the different particles and they are connected by an edge to another node if these two particles interact. This is straightforward to determine, since the WCA potential has a sharp interaction length. 
In addition, we assigned to each edge of the graph a weight according to a weight function $\zeta(r)$. We found $\zeta(r)=U_2(r)/\varepsilon$ (i.e., an edge weight corresponding to the energy of the represented interaction) to be a robust choice for outlining the region of clustering in $\mathrm{Pe}$-$\Phi$ space. Summing the edge weights for this choice of $\zeta$ then gives a dimensionless measure for the total potential energy $E_\mathrm{pot}$ stored in the system.
In the end, we averaged over all extracted data sets so that every point in the state diagram is based on $50$ graphs and therefore $50$ measurements of $E_\mathrm{pot}$. For the state diagrams, we considered the mean potential energy per particle, since this measure is independent of the size of the system. 

Using this procedure, one gets a clear and consistent measure for the state of a nonequilibrium system as an alternative to other common measures used for determining state diagrams of ABPs, such as visual inspection \cite{Stenhammar2014, RW}, the size of the biggest cluster \cite{Liao2018}, and the characteristic length \cite{StenhammarTAMC2013continuum,Stenhammar2014,BiniossekLVS2018,Jeggle2019, Broeker2019}. Our method has advantages compared to these other methods. Visual inspection has the drawback of not being objective and the size of the biggest cluster is a measure that depends on the size of the system. The characteristic length, albeit being dependent on the size of the system, works very well for ABPs where the clusters are rather static and will merge eventually into one big cluster. However, we observed that the clusters of circle swimmers behave much more dynamically: They rotate, break, and form again\footnote{In the supplementary material, we provide videos of MIPS occurring in systems of circle swimmers.}. Furthermore, the characteristic length becomes increasingly problematic at very high densities, since it cannot distinguish well between a random-close-packed system and an active cluster that has a similar density but an increased pressure in its interior. 
This comes from the fact that the interaction potential is very steep at low distances so that the mean potential energy per particle inside a MIPS cluster can be much larger than outside of such a cluster, while the particle densities inside and outside of a MIPS cluster are very similar. 
The graph-theoretical method can handle dynamical clusters and distinguish between those states, which makes it a perfect choice for the systems considered in the present work.

\section{\label{sec:Phasebehavior}Results} 
\begin{figure}[ht]
\centering
\includegraphics{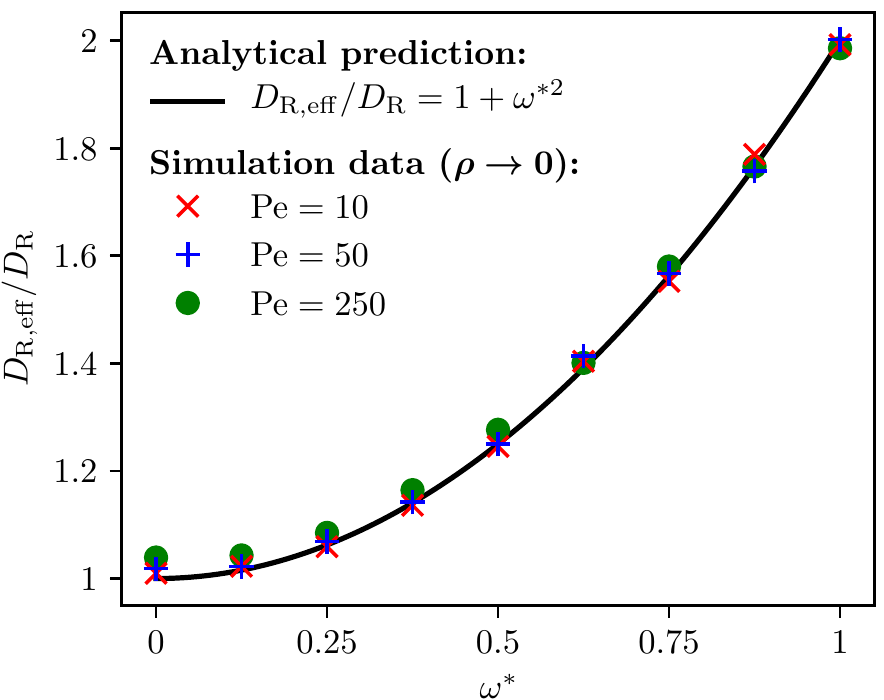}%
\caption{\label{fig:fig2}The dimensionless effective rotational diffusion $D_\mathrm{R, eff}/D_\mathrm{R}$ of circle swimmers as a function of the dimensionless angular propulsion velocity $\omega^*$ for different P\'eclet numbers $\mathrm{Pe}$. The solid line represents the analytical prediction \eqref{eqn:DReff}.}
\end{figure}

Before we investigate the collective dynamics of circle swimmers, we address their effective rotational diffusion $D_\mathrm{R, eff}$ and test the analytical result \eqref{eqn:DReff}. For this purpose, we study the long-time diffusion constant of a circle swimmer \cite{Teeffelen08,ten2009non}
\begin{equation}
D_\mathrm{cs} = \lim_{t\rightarrow\infty} \frac{1}{2 t} \big\langle (\vec{r}(t)-\vec{r}(0))^2\big\rangle.
\label{eq:DcsI}%
\end{equation}
For a noninteracting circle swimmer, we obtain from Eq.\ \eqref{eqn:2ndOrderModel} the exact analytical result 
\begin{equation}
D_\mathrm{cs} = D_\mathrm{T}\bigg(1+\frac{\mathrm{Pe}^2}{6\gamma}\bigg) 
\label{eq:DcsII}%
\end{equation}
with the rescaled effective rotational diffusion $\gamma = {D_\mathrm{R, eff}}/{D_\mathrm{R}} = 1+\omega^{*2}$. 
We therefore calculated $D_\mathrm{cs}$ using Eq.\ \eqref{eq:DcsI} based on Brownian dynamics simulations with $10,000$ noninteracting circle swimmers and determined the corresponding result for the rescaled effective rotational diffusion ${D_\mathrm{R, eff}}/{D_\mathrm{R}}$ by using Eq.\ \eqref{eq:DcsII}. This procedure was repeated for different values of $\omega^*$ and $\mathrm{Pe}$. 
In Fig.\ \ref{fig:fig2}, the simulation results for ${D_\mathrm{R, eff}}/{D_\mathrm{R}}$ are compared to the analytical result \eqref{eqn:DReff}.  
This figure shows that the rescaled effective rotational diffusion is independent of the P\'eclet number and growing with $\omega^{*2}$ and that the simulation results precisely confirm the analytical prediction \eqref{eqn:DReff}. 
The independence of $D_\mathrm{R, eff}/D_\mathrm{R}$ from $\mathrm{Pe}$ is reasonable, since $\mathrm{Pe}$ characterizes the activity corresponding to the translational propulsion of the particles, which does not affect their orientations, and the quadratic dependence on $\omega^{*}$ can be understood from the analogous scaling of the effective rotational temperature \eqref{eqn:TReff}. 
Since we analyze the rotational diffusion for noninteracting particles, the concept of an effective temperature is applicable both for the translational and for the rotational motion of the particles. In fact, in Eq.\ \eqref{eq:DcsII} a quadratic dependence on both the P\'eclet number $\mathrm{Pe}$ corresponding to the particles' translational motion and the rescaled angular propulsion velocity $\omega^*$ that constitutes a P\'eclet number corresponding to their rotational motion is observed.  
While the effective translational temperature \eqref{eqn:Teff} can no longer be defined when the interactions of the circle swimmers are taken into account, the effective rotational temperature \eqref{eqn:TReff} and the effective rotational diffusion coefficient \eqref{eqn:DReff} are still applicable in this case, since the considered spherical particles have only positional but not angular interactions. 
If also angular interactions come into play (e.g., by investigating anisotropic particles like ellipsoids), we suspect $D_\mathrm{R, eff}$ to become density-dependent.

\begin{figure*}[!t]
\centering
\includegraphics{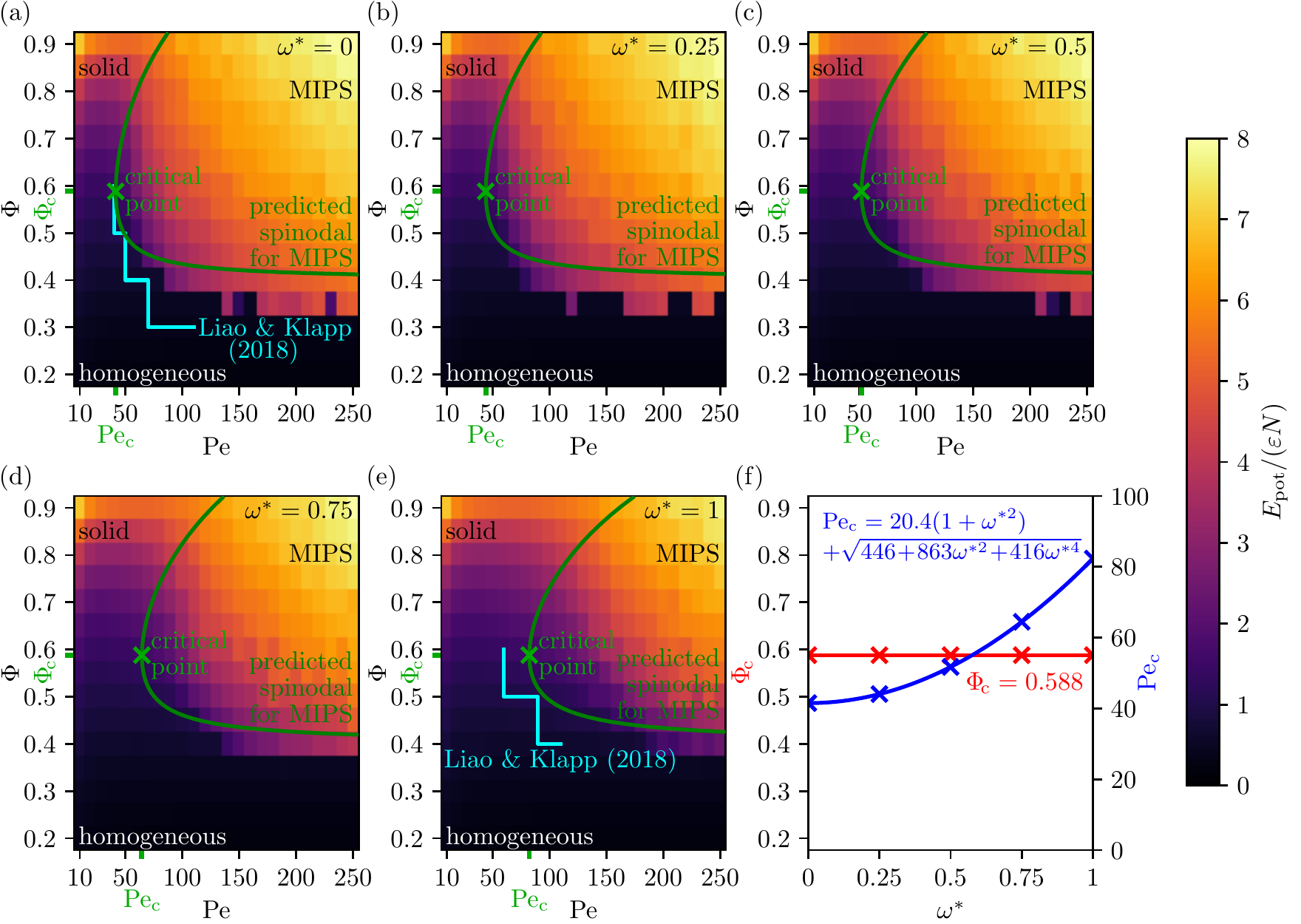}%
\caption{\label{fig:fig3}State diagram for (a) ABPs and (b)-(e) circle swimmers with various rescaled angular velocities $\omega^*$. The color bar shows the rescaled mean potential energy per particle $E_\mathrm{pot}/(\varepsilon N)$ that reveals the different states of the system. These include a homogeneous state at low packing densities $\Phi$, a solid state at very high $\Phi$ and low P\'eclet numbers $\mathrm{Pe}$, and a state corresponding to MIPS at high $\Phi$ and $\mathrm{Pe}$.
In subfigures (a) and (e), the computer-simulation results of Ref.\ {[\hspace*{-.7 ex}\citenum{Liao2018}]} for the spinodal are shown for comparison.
The associated curves in these subfigures correspond to the criterion that the fraction of the largest cluster passes a value of about $0.3$. 
Our analytical prediction for the spinodal describing the onset of MIPS (see Eq.\ \eqref{eqn:SpinCon}) is indicated in each state diagram. 
Also our analytical predictions for the coordinates of the critical point, being the critical packing density $\Phi_\mathrm{c}$ and the critical P\'eclet number $\mathrm{Pe}_\mathrm{c}$, are marked in the state diagrams. 
Subfigure (f) shows the dependence of $\Phi_\mathrm{c}$ and $\mathrm{Pe}_\mathrm{c}$ on $\omega^*$.}
\end{figure*}

Furthermore, we investigated the collective dynamics of interacting circle swimmers. For this purpose, we first applied a linear-stability analysis to our $2$nd-order-derivatives model \eqref{eqn:2ndOrderModel}. This analysis yields the spinodal condition describing the onset of MIPS
\begin{equation}
D(\rho) = 0. 
\label{eqn:SpinCon}%
\end{equation}
Figure \ref{fig:fig3} presents our simulation results for the state of a system of circle swimmers as a function of the P\'eclet number $\mathrm{Pe}$ and packing density $\Phi$ in the form of state diagrams for various values of the rescaled angular velocity $\omega^*$ as well as our predictions for the spinodal given by Eq.\ \eqref{eqn:SpinCon}. 
For comparison, also the earlier simulation results for the spinodal, which are presented in Ref.\ \mbox{\onlinecite{Liao2018}} for $\omega^*=0$ and $\omega^*=1$, are plotted (see Figs.\ \ref{fig:fig3}(a) and \ref{fig:fig3}(e)). 
These results are consistent with ours, which cover a larger parameter space and have a finer resolution.
Our simulation results show three different states. 
They are a homogeneous state occurring at low densities, a solid state at high densities and low P\'eclet numbers, and a MIPS state at high densities and high P\'eclet numbers. 
Interestingly, for increasing rescaled angular velocity $\omega^*$, the MIPS region is shifted to higher P\'eclet numbers and packing densities.
However, the maximum value for the mean potential energy per particle does not change significantly in the cluster state when $\omega^*$ is increased. 
The solid region is similar to the findings of Refs.\ \onlinecite{Digregorio2018,vanDamme19} and seems to be only weakly affected by the additional torque $\omega^*$.   
Our analytical prediction \eqref{eqn:SpinCon} for the spinodal corresponding to the onset of MIPS is in good agreement with our simulation results for the state diagram and those of Ref.\ \onlinecite{Liao2018}. 
It is important to take into account that the simulation data can show MIPS slightly outside the predicted spinodal, since fluctuations in the simulations can lead to a transition into the cluster state if the system is in the binodal region. 
Furthermore, the results shown in Fig.\ \ref{fig:fig3} are in line with the predictions of Ref.\ \onlinecite{Navarro_2015}, where they studied ABPs with different rotational diffusion constants. They only observed MIPS in systems of small angular reorientation but not in systems corresponding to a high angular diffusion (see Fig.\ 8 in Ref.\ \onlinecite{Navarro_2015}). 

With the equation \eqref{eqn:SpinCon} for the spinodal, we were also able to derive an analytical prediction for the associated critical point, which is given by the critical P\'eclet number $\mathrm{Pe}_\mathrm{c}$ and the critical packing density $\Phi_\mathrm{c}$. Our predictions for $\mathrm{Pe}_\mathrm{c}$ and $\Phi_\mathrm{c}$ as functions of $\omega^*$ are given by   
\begin{align}
\mathrm{Pe}_\mathrm{c} &= 20.4(1+\omega^{*2})+\sqrt{446 + 863 \omega^{*2} + 416 \omega^{*4}},\label{eqn:Pec}\\%
\Phi_\mathrm{c} &= 0.588 .
\end{align}
These expressions are visualised in Fig.\ \ref{fig:fig3}(f).
Interestingly, for increasing $\omega^*$, the critical point shifts to higher $\mathrm{Pe}_\mathrm{c}$ but remains at a constant $\Phi_\mathrm{c}$. 
To the best of our knowledge, the critical point for systems of circle swimmers has not yet been located in the literature.

\section{\label{sec:Conclusions}Conclusions}
We have investigated the dynamics of interacting spherical Brownian circle swimmers in two spatial dimensions based on deriving a predictive field theory describing their collective behavior and performing Brownian dynamics simulations. 
The derivation yielded a $2$nd-order-derivatives model and a $4$th-order-derivatives model that are applicable for small values of the particles' rescaled angular velocity $\omega^*$. Using the $2$nd-order model, we showed that the dynamics of circle swimmers can be mapped onto that of common ABPs via introducing an effective rotational diffusion coefficient $D_\mathrm{R, eff}$. 
We also obtained analytical expressions for the spinodal corresponding to the onset of MIPS and for the associated critical point, which allow to assess how the spinodal and the critical point depend on $\omega^*$. 
All these analytical predictions were found to be in good agreement with our simulation results and available results from the literature \cite{Liao2018}. To determine state diagrams where MIPS can clearly be identified from our simulation data, we computed the potential energy of the system by means of a weighted graph-theoretical network. 
Using these analytical and numerical methods, we also investigated the suppression of MIPS by circle swimming, which was previously reported in Ref.\ \onlinecite{Liao2018}, in more detail.  
We found that the spinodal shifts to larger P\'eclet numbers $\mathrm{Pe}$ and packing densities $\Phi$ when $\omega^*$ is increased, whereas the associated critical point shifts only to larger $\mathrm{Pe}$ but remains at a constant $\Phi$.   
A solid state, which was also observed in the state diagrams, showed no significant dependence on $\omega^*$.  
These results show that the occurrence of MIPS and thus the collective dynamics of circle swimmers can be tuned via their angular velocity. Since circle swimming can be induced and easily tuned, e.g., by a rotating external magnetic field in combination with magnetic ABPs, it offers a simpler way to suppress MIPS than alignment \cite{Barre2015,vanDamme19}.  
This makes suspensions of circle-swimming active particles to a particularly interesting class of active materials.  

The $4$th-order-derivatives model we derived can be used for further investigations of the collective behavior of interacting circle swimmers. With this model, one could address, e.g., the occurring finite cluster sizes or interfacial profiles. 
For example, the model could be used to study at which value of the particles' rescaled angular velocity $\omega^*$ the complete phase separation occurring in systems of ordinary ABPs is replaced by the arrested phase separation that can be observed for circle swimmers. 
Furthermore, we believe that our method for computing the potential energy of the system via a graph-theoretical network will prove useful also for other dynamical many-particle systems. Using a weighted graph-theoretical network can have significant benefits, since such networks are mathematically well established and numerous further parameters that are associated to a graph might turn out to be interesting properties for characterizing a physical system. 
In the future, this study could be extended towards circle swimmers with other shapes, which will very likely lead to the observation of fascinating new effects.

\appendix
\section{\label{app:Coefficients}Appendix}
We introduce the angular relaxation time $\tau = 1/D_\mathrm{R}$ for abbreviation. The coefficients for the first-order corrections in $\omega$ occurring in Eq.\ \eqref{eqn:4thOrderModel} are then given by the following expressions:
\begin{align}
    \begin{split}
        \mu_1 &= -\frac{\tau ^2 v_0}{32 \pi } (64 D_\mathrm{T} \tau  (A(0,1,-1)+A(0,1,0))\\
        &\quad\, +\tau ^2 v_0^2 (-43 A(0,1,-1)+15 A(0,1,0)+A(0,1,1))\\
        &\quad\, -\tau  v_0 (16 A(1,0,-1)-32 A(1,0,0)+16 A(1,0,1)\\
        &\quad\, +3 A(1,2,-2)-16 A(1,2,-1)+3 A(1,2,0))\\
        &\quad\, +4 (A(2,1,-1)-A(2,1,0))),
    \end{split}\\
    \begin{split}
        \mu_2 &= \frac{\tau ^2}{32 \pi ^2} (128 D_\mathrm{T} \tau  A(0,1,0) (A(0,1,-1)+A(0,1,0))\\
        &\quad\, +\tau ^2 v_0^2 (-48 A(0,1,-1)^2+5 (A(0,1,1)-42 A(0,1,0))\\
        &\quad\, \quad\, A(0,1,-1)+60 A(0,1,0)^2+A(0,1,1)^2\\
        &\quad\, +20 A(0,1,0) A(0,1,1))-\tau  v_0 (A(0,1,1) (3 A(1,2,-2)\\
        &\quad\, +2 A(1,2,0))+A(0,1,-1) (32 A(1,0,-1)-32 A(1,0,0)\\
        &\quad\, +32 A(1,0,1)+3 A(1,2,0))+A(0,1,0) (64 A(1,0,-1)\\
        &\quad\, -96 A(1,0,0)+64 A(1,0,1)+9 A(1,2,-2)\\
        &\quad\, -64 A(1,2,-1)+18 A(1,2,0)))+A(1,2,-2) A(1,2,0)\\
        &\quad\, +8 A(0,1,0) A(2,1,-1)+20 A(0,1,-1) A(2,1,0)\\
        &\quad\, +A(1,2,0)^2-4 A(0,1,0) A(2,1,0)),
    \end{split}\\
    \begin{split}
        \mu_3 &= -\frac{\tau ^3}{24 \pi ^3} A(0,1,0) (\tau  v_0 (-192 A(0,1,-1)^2+(25 A(0,1,1)\\
        &\quad\, -329 A(0,1,0)) A(0,1,-1)+75 A(0,1,0)^2+4 A(0,1,1)^2\\
        &\quad\, +53 A(0,1,0) A(0,1,1))-64 A(0,1,0) A(1,0,-1)\\
        &\quad\, +64 A(0,1,0) A(1,0,0)-64 A(0,1,0) A(1,0,1)\\
        &\quad\, -6 A(0,1,0) A(1,2,-2)-6 A(0,1,1) A(1,2,-2)\\
        &\quad\, +64 A(0,1,0) A(1,2,-1)-15 A(0,1,0) A(1,2,0)\\
        &\quad\, -4 A(0,1,1) A(1,2,0)-A(0,1,-1) (64 A(1,0,-1)\\
        &\quad\, -64 A(1,0,0)+64 A(1,0,1)+9 A(1,2,0))),
    \end{split}\\
    \begin{split}
        \mu_4 &= \frac{\tau ^4}{8 \pi ^4} A(0,1,0)^2 (-96 A(0,1,-1)^2+(15 A(0,1,1)\\
        &\quad\, -81 A(0,1,0)) A(0,1,-1)+15 A(0,1,0)^2+2 A(0,1,1)^2\\
        &\quad\, +17 A(0,1,0) A(0,1,1)),
    \end{split}\\
    \begin{split}
        \mu_5 &= -\frac{\tau ^2}{16 \pi ^2} (128 D_\mathrm{T} \tau  A(0,1,0) (A(0,1,-1)+A(0,1,0))\\
        &\quad\, +\tau ^2 v_0^2 (8 A(0,1,-1)^2+(13 A(0,1,1)-68 A(0,1,0)) \\
        &\quad\,\quad\,  A(0,1,-1)+5 A(0,1,0) (7 A(0,1,0)+4 A(0,1,1)))\\
        &\quad\, -\tau  v_0 (3 A(0,1,1) A(1,2,-2)+A(0,1,-1) (8 A(1,0,-1)\\
        &\quad\, +16 A(1,0,0)+8 A(1,0,1)+3 A(1,2,0))\\
        &\quad\, +A(0,1,0) (24 A(1,0,-1)-16 A(1,0,0)+24 A(1,0,1)\\
        &\quad\, +9 A(1,2,-2)-32 A(1,2,-1)+6 A(1,2,0)))\\
        &\quad\, +A(1,2,-2) A(1,2,0)+8 A(0,1,0) A(2,1,-1)\\
        &\quad\, +4 A(0,1,-1) A(2,1,0)-4 A(0,1,0) A(2,1,0)),
    \end{split}\\
    \begin{split}
        \mu_6 &= -\frac{\tau^3}{8 \pi^3} A(0,1,0) (\tau  v_0 (64 A(0,1,-1)^2+(191 A(0,1,0)\\
        &\quad\, -67 A(0,1,1)) A(0,1,-1)-85 A(0,1,0)^2-2 A(0,1,1)^2\\
        &\quad\, -93 A(0,1,0) A(0,1,1))+64 A(0,1,0) A(1,0,-1)\\
        &\quad\, +64 A(0,1,0) A(1,0,1)+12 A(0,1,0) A(1,2,-2)\\
        &\quad\, +12 A(0,1,1) A(1,2,-2)-64 A(0,1,0) A(1,2,-1)\\
        &\quad\, +9 A(0,1,0) A(1,2,0)-2 A(0,1,1) A(1,2,0)\\
        &\quad\, +A(0,1,-1) (64 A(1,0,-1)+64 A(1,0,1)+9 A(1,2,0))),
    \end{split}\raisetag{20ex}\\
    \begin{split}
        \mu_7 &= -\frac{\tau^4}{2 \pi^4} A(0,1,0)^2 (-80 A(0,1,-1)^2+(41 A(0,1,1)\\
        &\quad\, -55 A(0,1,0)) A(0,1,-1)+25 A(0,1,0)^2+2 A(0,1,1)^2\\
        &\quad\, +43 A(0,1,0) A(0,1,1)).
    \end{split}
\end{align}

\section*{Conflicts of interest}
There are no conflicts to declare.

\section*{Acknowledgements}
We thank Andreas Menzel for helpful discussions.
R.W.\ is funded by the Deutsche Forschungsgemeinschaft (DFG, German Research Foundation) -- WI 4170/3-1. 
The simulations for this work were performed on the computer cluster PALMA II of the University of M\"unster.

\balance
\bibliography{refs} 
\bibliographystyle{rsc} 

\end{document}